# A PEMFC-based combined cooling heating and power (CCHP) system with flexible energy supply: Thermodynamic and economic analyses

Yuanming Wang[1], Shaowen Deng[1], Rui Chen[1,2], Zhen Zeng[1,2], Tianyou Wang[1,2], Zhizhao Che[1,2,*]

1. State Key Laboratory of Engines, Tianjin University, Tianjin, 300350, China

2. National Industry-Education Platform of Energy Storage, Tianjin University, Tianjin, 300350, China

*Corresponding author, chezhizhao@tju.edu.cn

## Abstract

Proton exchange membrane fuel cell (PEMFC) systems offer a key approach to hydrogen utilization, and PEMFC-based combined cooling, heating, and power (CCHP) systems pave the way for an efficient and clean energy supply to buildings. In conventional PEMFC-CCHP systems, the heating/cooling capacity and electrical power output are strongly coupled, making it difficult to meet diverse energy demands. This paper presents a novel energy system that integrates an organic Rankine cycle and an absorption heat pump in a parallel configuration, which enables flexible regulation of electricity-cooling capacities in summer and electricity-heating capacities in winter by adjusting the splitting ratio of the waste heat. The impacts of the splitting ratio and key operating parameters on thermodynamic performance and economic performance are quantitatively evaluated. The results show that the ORC can improve electrical efficiency by 2.19 percentage points in summer and 2.78 percentage points in winter. When the current density is fixed at 0.4 A cm$^{-2}$ and the splitting ratio increases from 0 to 0.5, the cooling capacity of the system varies from 1294 to 647 W, and the heating capacity varies from 2660 to 1330 W. The economic performance is more sensitive to electricity price and hydrogen price than to other parameters, confirmed by their high sensitivity coefficients for net present value (NPV) and internal rate of return (IRR). This system possesses excellent thermodynamic and economic properties, thereby offering significant potential for reducing building energy consumption and carbon emissions.

## 1. Introduction

High requirements have been placed on energy conservation and carbon reduction to cope with globalized climate change challenges. Hydrogen energy has become a vital choice for its zero-carbon characteristics [1]. Hydrogen-fueled proton exchange membrane fuel cells (PEMFCs) demonstrate superior performance characteristics, including exceptional energy conversion efficiency, zero-emission operation, and low noise, hence providing a vital route for energy saving and carbon reduction [2-5]. An important area of PEMFC application is the building sector, which contributes 36% of all greenhouse gases emitted worldwide and consumes 40% of the total energy consumption [6]. PEMFCs directly convert chemical energy into electrical energy, while about half of the chemical energy escapes as heat [7, 8], and the temperature control of the PEMFC should be within the normal range for proper functioning [9, 10]. Therefore, the development of PEMFC waste heat reutilization systems for building energy supply has important application prospects [11-13].

Common methods of PEMFC waste heat utilization include PEMFC combined heat and power (CHP) systems [14-16], PEMFC combined cooling, heating, and power (CCHP) systems [17-21], and PEMFC combined cooling and power (CCP) systems [22, 23]. Ning et al. [16] investigated the electrical and thermal performance of a PEMFC-CHP system across various operational modes for different application scenarios. Xu et al. [22] built a PEMFC-CCP system with an efficient adsorption chiller to enhance efficiency by efficiently reusing waste heat. Lu et al. [18] proposed a PEMEC-CCHP system, and a solar thermal collector was used to increase the waste heat grade, thereby improving the cooling/heating flexibility supply. Liang et al. [24] developed a 1000-kWe PEMFC-CCHP system that is integrated with a "methanol-solar-to-X" system to reduce environmental pollution and minimize annual costs. Zhao et al. [25] designed a PEMFC-CCHP system incorporating dehumidification, along with economic evaluation and dynamic responses studies. The result shows that the system, combined with a parabolic trough solar collector, can reduce hydrogen consumption. Zhang et al. [26] developed an efficient PEMFC-CCHP system that is coupled with thermochemical energy storage and solar energy, which has high application value and development potential in power generation, energy storage, cooling, and heating. To accommodate variable dynamic demand, energy storage devices are integrated into the system to achieve peak shaving and valley filling, thereby improving the alignment between energy supply and demand. Typically, a water tank serves as thermal energy storage [7], while batteries or supercapacitors are employed for electrical energy storage [27, 28].

Conventional waste heat utilization for PEMFC systems typically incorporates PEMFC systems with absorption systems, organic Rankine cycle systems, and other components to form various cycles. However, these systems exhibit inherent operational constraints: when the heat recovery is fixed, the energy supply is also uniquely determined. The traditional system with limited regulation capacity significantly restricts operational flexibility in waste heat utilization, inevitably leading to energy losses under variable load conditions. To adapt to the varying electricity and cooling/heat load, Ma et al. [29] proposed an absorption cooling/heating cogeneration cycle for high-temperature PEMFC that decouples the electricity and heating/cooling supply through distributing the vapor ratio to the turbine branch and evaporator branch. However, this approach is only applicable to high-temperature fuel cells. It is needed to explore low-temperature power generation methods for low-temperature fuel cells.

Converting excess waste heat into electricity can significantly enhance the exergy efficiency while reducing the dependence on energy storage. However, the low-grade thermal energy from PEMFC stacks poses a significant challenge for thermoelectric conversion. The Organic Rankine Cycle (ORC) is highly effective at recovering low-grade thermal energy and converting it to electricity [30, 31]. Lu et al. [14] developed an integrated PEMFC-CHP-ORC configuration, demonstrating 5.1% and 4.3% enhancements in exergy efficiency and electrical efficiency over conventional systems, respectively. Liu et al. [32] proposed a configuration that uses an organic working fluid as a cooling medium directly and then converts the thermal energy to electricity by an ORC system. Mohammadkhani et al. [33] proposed a multigeneration system and identified the SOFC, HRSG, and MED-TVC units as the primary sources of exergy destruction through a comprehensive 4E analysis.

The economics of energy systems are a key factor in determining their viability. In the past, many scholars have analyzed the economics of the system to determine its application value and feasibility, and the economic analysis is mainly based on key parameters such as the dynamic payback period (DPP), net present value (NPV), levelized cost of exergy (LCOE)[34], and internal rate of return (IRR) [35]. In addition, tornado diagrams and sensitivity factors serve as useful tools in sensitivity analysis to quantify the influence of the input parameter uncertainties on the results [36]. El-Temtamy et al. [35] evaluated seven upgrading schemes for the upgrading of atmospheric residues economically, identified the most profitable option, and conducted sensitivity analyses by tornado diagrams and spider charts to assess its financial robustness. Ma et al. [4] proposed a hybrid system by combining high-temperature PEMFC and ammonia absorption power generation, demonstrating favorable economic viability, with thermodynamic evaluation showing the system achieves payback within 5.2 years of its 20-year service life. Zhao et al. [37] developed a PEMFC trigeneration configuration, while the operation parameters on the 4E (environment, economy, exergy, and energy) are discussed and optimized. An integrated residential PEMFC-CCHP system was developed [38] with multi-objective performance assessment and evolutionary algorithm-driven optimization. Lombardo et al. [39] carried out a solar-driven CCHP system that mainly includes a solar thermal collector, photovoltaic plant, and micro-ORC system to support a nearly zero energy building. The analysis yielded a 6-year median payback period, and the corresponding median net present value (NPV) is 50 thousand euros, suggesting outstanding market potential.

Unlike the existing PEMFC–CCHP and PEMFC–ORC systems, this study proposes a novel PEMFC residual heat utilization configuration that combines an absorption heat pump (AHP) with an organic Rankine cycle (ORC) in a parallel coupling arrangement. By adjusting the splitting ratio of the waste heat, the system enables flexible regulation of electricity-cooling capacities in summer and electricity-heating capacities in winter, thereby achieving decoupling of electricity generation, cooling, and heating. When the demand for electrical power is high while the demand for cooling/heating is low, waste heat is preferentially routed to the ORC branch to enhance electrical output; conversely, waste heat is directed primarily to the AHP branch to increase cooling/heating capacity, thereby meeting the comprehensive energy demands in various scenarios.

## 2. System modeling

Figure 1 schematically shows the PEMFC-CCHP system setup. The PEMFC-CCHP system includes a PEMFC system, an AHP system, and an ORC system. The PEMFC system includes an anode circuit (states 1-5) that realizes the circulation of hydrogen by an injector, a cathode circuit (states 6-11), and a cooling medium circuit (states 26-33). Deionized water is used as the cooling medium of the fuel cell. A pump drives the circulation of deionized water to achieve the cooling of the fuel cell and the recovery of waste heat (states 26-33). The ORC system (states 12-15) and the AHP (states 16-25) system are arranged in parallel. The heated deionized water is divided into two channels after passing through a three-way valve. By adjusting the opening degree of the three-way valve, deionized water was flexibly distributed to the AHP and ORC system (states 29-32). In the ORC system, the vapor generated in the evaporator expands through a turbine to generate electricity (states 12-13), after which it is condensed in the condenser (states 13-14). The condensed organic working fluid is then pumped back to the evaporator, thereby completing the thermodynamic cycle (states 14-15). In the AHP system, waste heat is utilized in the generator to desorb the refrigerant from the lithium bromide solution. The desorbed refrigerant vapor then releases heat in the condenser (states 16-17), undergoes throttling in the expansion valve, and evaporates in the evaporator (states 17-19). Finally, it is reabsorbed by the lithium bromide solution in the absorber. During heating operation, the cooling water first passes through the absorber for preheating and then enters the condenser for further heat upgrading, forming a serial configuration that provides heating capacity to the user [40]. Driven by the heated deionized water, the AHP generates a cooling effect in summer or a heating effect in winter, while the ORC system generates electricity by a turbine. The diagram focuses on key components, and auxiliary equipment deemed non-essential to the thermodynamic analysis has been omitted.

In this study, the ambient temperature is 303.15 K in summer and 278.15 K in winter, and the ambient pressure is 1 bar.

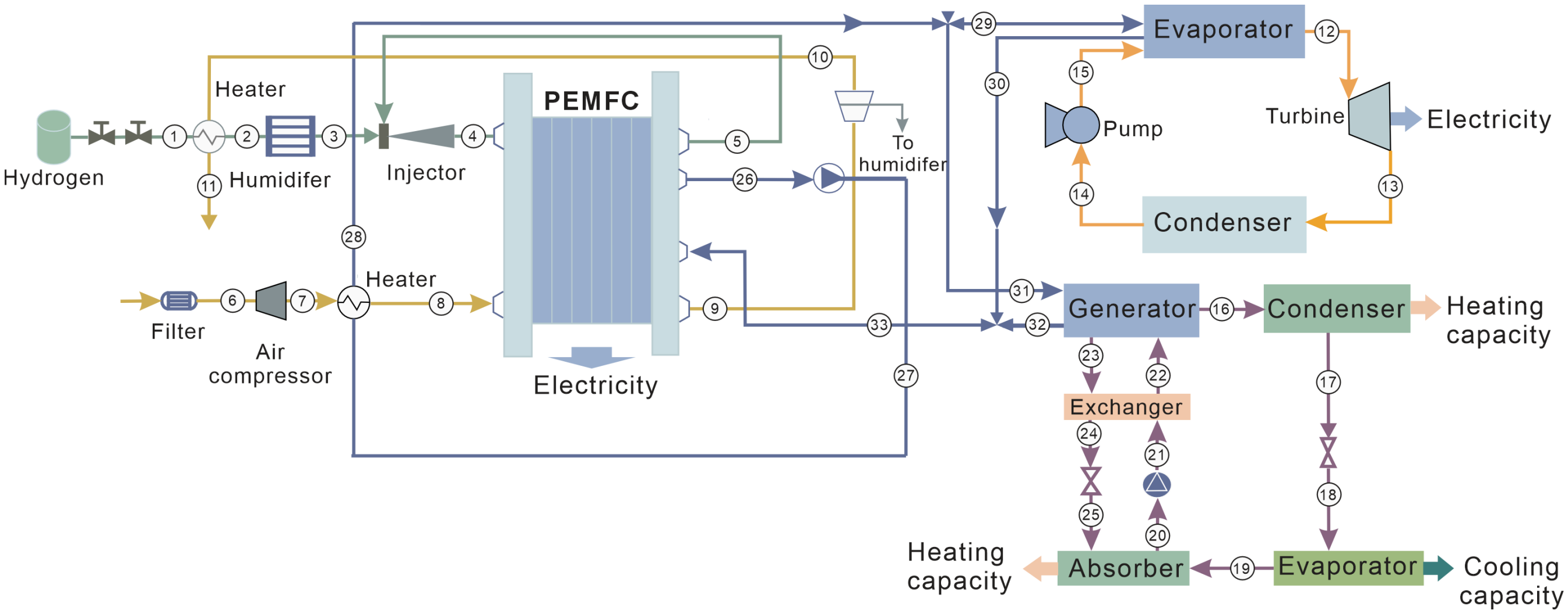

Figure 1: Configuration layout of the integrated PEMFC-based CCHP system.

### 2.1 PEMFC stack and the air compressor

For PEMFC system modeling, the following simplifications are implemented: (1) Unless otherwise indicated, thermal losses and pressure drops of the component are ignored. (2) The stack exhibits uniform operational conditions across all cells. (3) Air and hydrogen are treated as ideal gases. (4) The anode inlet hydrogen is pure. (2) Air consists of 21% oxygen and 79% nitrogen.

The PEMFC voltage $E_{\mathrm{cell}}$ results from subtracting all overpotentials from the Nernst potential, and the Nernst potential $E_{\mathrm{Nernst}}$ is calculated as,

$$E_{\mathrm{Nernst}} = 1.229 - 8.456\times10^{-4}\left(T - 298.15\right) + 4.3085\times10^{-5} T \ln\left(P_{\mathrm{H_2}} P_{\mathrm{O_2}}^{0.5}\right) \tag{1}$$

where $T$ represents the PEMFC temperature, and $P$ stands for pressure ($10^5$ Pa). The value of the ideal gas constant $R$ is 8.314 J mol$^{-1}$ K. The irreversible voltage losses include activation overpotential $E_{\mathrm{act}}$, ohmic overpotential $E_{\mathrm{ohm}}$, and concentration overpotential $E_{\mathrm{con}}$, which can be calculated as [41],

$$\begin{aligned} E_{\mathrm{act}} = 0.9512 - &\left[0.00286 + 0.0002\ln(A) + 4.3\times10^{-5}\ln(\ell_{\mathrm{H_2}})\right]T \\ &-7.8\times10^{-5} T\ln(\ell_{\mathrm{O_2}}) + 1.98\times10^{-4} T\ln(I) \end{aligned} \tag{2}$$

$$\ell_{\mathrm{H_2}} = \frac{P_{\mathrm{H_2}}}{1.09\times10^{6}\exp\left(77/T\right)} \tag{3}$$

$$\ell_{\mathrm{O_2}} = \frac{P_{\mathrm{O_2}}}{5.08\times10^{6}\exp\left(-498/T\right)} \tag{4}$$

$$E_{\mathrm{ohm}} = I\left(\frac{r_{\mathrm{m}}\sigma_{\mathrm{m}}}{A} + r_{\mathrm{elec}}\right) \tag{5}$$

$$r_{\mathrm{m}} = \frac{181.6\times\left[0.062\times(T/303)^{2} j^{2.5} + 1 + 0.03j\right]}{(\lambda - 0.634 - 3j)\exp(4.18\times\frac{T-303}{T})} \tag{6}$$

$$\lambda = \begin{cases} 0.043 + 17.81a_{\mathrm{W}} - 39.85{a_{\mathrm{W}}}^{2} + 36{a_{\mathrm{W}}}^{3}, & 0 < a_{\mathrm{W}} \le 1 \\ 1.4 + 1.4(a_{\mathrm{W}} - 1), & 1 < a_{\mathrm{W}} \le 3 \end{cases} \tag{7}$$

$$E_{\mathrm{con}} = -\frac{RT}{\kappa F}\ln(1 - \frac{j}{j_{1}}) \tag{8}$$

$$E_{\mathrm{cell}} = E_{\mathrm{nernst}} - E_{\mathrm{act}} - E_{\mathrm{ohm}} - E_{\mathrm{con}} \tag{9}$$

where A is the active area (cm$^2$), $I$ stands for the current (A), $\sigma_{\mathrm{m}}$ represents the membrane thickness (cm), $r_{\mathrm{m}}$ is the membrane specific resistivity for the flow of hydrated protons (Ω cm), $r_{\mathrm{elec}}$ is the contact resistance (Ω), $j$ is the current density (A cm$^{-2}$), $F$ is the Faraday constant (C mol$^{-1}$), $\kappa$ is the number of electrons transferred in the electrochemical reaction, $\lambda$ is the effective membrane water content, and $a_{\mathrm{W}}$ is the relative humidity.

The stack voltage $E_{\mathrm{st}}$ and stack power $W_{\mathrm{st}}$ are defined as,

$$E_{\mathrm{st}} = N_{\mathrm{cell}} E_{\mathrm{cell}} \tag{10}$$

$$W_{\mathrm{st}} = E_{\mathrm{st}} I \tag{11}$$

where $N_{\mathrm{cell}}$ is the number of the PFMFC. $Q_{\mathrm{st}}$ is the heat produced by the PEMFC stack (Eq. (12)), and the heat removed by the gas is efficiently calculated as Eq. (13).

$$Q_{\mathrm{st}} = N_{\mathrm{cell}} I (1.48 - E_{\mathrm{cell}}) \tag{12}$$

$$\begin{aligned} Q_{\mathrm{gas}} &= c_{\mathrm{O_2}} \left[ 0.21 (\beta_{\mathrm{ca}} - 1) n_{\mathrm{air}} T - 0.21 n_{\mathrm{air}} T_{\mathrm{in}} \right] + c_{\mathrm{N_2}} \left[ 0.79 n_{\mathrm{air}} (T - T_{\mathrm{in}}) \right] \\ &+ c_{\mathrm{H_2O}} \left[ T \left( \frac{n_{\mathrm{H_2}}}{\beta_{\mathrm{an}}} + n_{\mathrm{H_2O}}^{\mathrm{ca}} \right) - T_{\mathrm{in}} n_{\mathrm{H_2O}}^{\mathrm{ca}} \right] \end{aligned} \tag{13}$$

$$n_{\mathrm{H_2}} = \beta_{\mathrm{an}} \frac{N_{\mathrm{cell}} I}{2F} \tag{14}$$

$$n_{\mathrm{air}} = \frac{n_{\mathrm{O_2}}}{\mathbb{R}_{\mathrm{O_2}}} = \beta_{\mathrm{ca}} \frac{N_{\mathrm{cell}} I}{4 \mathbb{R}_{\mathrm{O_2}} F} \tag{15}$$

where $n$ is the molar flow rate (mol s$^{-1}$), and $\mathbb{R}_{\mathrm{O_2}}$ represents the molar fraction of oxygen in atmospheric air. $T_{\mathrm{in}}$ is the temperature of the stack inlet gas, and is 5 K cooler than the stack operation temperature. The air compressor is used to pressurize the air to meet the inlet requirements of the PEMFC stack. The process of gas compression is modeled as an isentropic process, with the air compressor power consumption given by Eq. (16).

$$W_{\mathrm{com}} = c_{\mathrm{air}} n_{\mathrm{air}} \frac{T_0}{\eta_{\mathrm{com}}} \left[ \left( P_{\mathrm{com}} / P_0 \right)^{\frac{\gamma - 1}{\gamma}} - 1 \right] \tag{16}$$

where $P_{\mathrm{com}}$ and $P_0$ represent the stack inlet and ambient pressures, respectively, $\eta_{\mathrm{com}}$ is the air compressor efficiency, and its value is 0.78 [32]. The compression process within an air compressor generates a large amount of heat that is defined in Eq. (17), and the power consumed by the cooling water pump is defined in Eq. (18).

$$Q_{\mathrm{com}} = c_{\mathrm{air}} n_{\mathrm{air}} \left[ \frac{T_0}{\eta_{\mathrm{com}}} \left( \left( P_{\mathrm{com}} / P_0 \right)^{\frac{\gamma - 1}{\gamma}} - 1 \right) + T_0 - T_{\mathrm{in}} \right] \tag{17}$$

$$W_{\mathrm{p}} = \frac{\Re \rho g H q_{\mathrm{v}}}{\eta_{\mathrm{p}}} \tag{18}$$

where $c$ represents the specific heat capacity (J mol$^{-1}$ K$^{-1}$), $\eta_{\mathrm{p}}$ represents the pump efficiency(-), $\rho$ stands for the density (kg m$^{-3}$), $H$ represents the lift (m). $\Re$ stands for the abundance factor (-), which is essentially an empirical coefficient used to correct the deviation between the actual operating conditions and the theoretical design conditions. $q_{\mathrm{v}}$ corresponds to the volume flow rate of the cooling water (m$^3$ s$^{-1}$). The recovered heat and the net electrical power of the PEMFC stack are defined by Eqs. (19) and (20), respectively.

$$Q = Q_{\mathrm{st}} + Q_{\mathrm{com}} - Q_{\mathrm{gas}} \tag{19}$$

$$W_{\mathrm{st,net}} = W_{\mathrm{st}} - W_{\mathrm{com}} - W_{\mathrm{p}} \tag{20}$$

**2.2 Absorption heat pump**

In the PEMFC-CCHP system, waste heat drives a LiBr/$H_2O$ AHP to generate a cooling effect in summer or a heating effect in winter. In summer, the condenser temperature, absorber temperature, and evaporator temperature are 303.15, 298.15, and 276.15 K, respectively. In winter, the condenser temperature, absorber temperature, and evaporator temperature are 325.15, 298.15, and 283.15 K, respectively. The stack temperature decides the generator temperature, and the effectiveness of the heat exchanger is 0.707. The AHP model relies on these key assumptions. (1) The pressure loss in the piping is ignored. (2) The system operates in steady state. (3) The refrigerant leaving the condenser and evaporator is saturated. (4) The solution leaving the generator and absorber is saturated. (5) The throttling process is isenthalpic. The mathematical model of an AHP primarily relies on the conservation of mass, concentration, and energy, which can be defined as:

$$\sum m_{\text{in}} - \sum m_{\text{out}} = 0 \tag{21}$$

$$\sum (mx)_{\text{in}} - \sum (mx)_{\text{out}} = 0 \tag{22}$$

$$\sum (mh)_{\text{in}} - \sum (mh)_{\text{out}} + Q + W = 0 \tag{23}$$

The balance equation in the generator can be given as follows:

$$m_{22}h_{22} - m_{23}h_{23} - m_{16}h_{16} + Q_{\text{g}} = 0 \tag{24}$$

$$Q_{\text{g}} = m_{31}(h_{31} - h_{32}) \tag{25}$$

$$m_{22}x_{22} = m_{23}x_{23} \tag{26}$$

The balance equation in the condenser is

$$Q_{\text{c}} = m_{16}(h_{16} - h_{17}) \tag{27}$$

The balance equation in the evaporator can be defined as

$$Q_{\text{e}} = m_{18}(h_{19} - h_{18}) \tag{28}$$

The balance equation in the absorber is

$$m_{20}h_{20} - m_{19}h_{19} - m_{25}h_{25} + Q_{\text{a}} = 0 \tag{29}$$

$$m_{20}x_{20} = m_{25}x_{25} \tag{30}$$

In summer, the cooling capacity is provided by the evaporator. In winter, the heating capacity provided by the AHP is the sum of the heat released in the absorber (exothermic absorption of refrigerant vapor into the solution) and the heat released in the condenser (condensation of refrigerant vapor). The cooling capacity $Q_{\text{C}}$ and the heating capacity $Q_{\text{H}}$ of the AHP are:

$$Q_{\text{C}} = Q_{\text{e}} = m_{18}(h_{19} - h_{18}) \tag{31}$$

$$Q_{\text{H}} = Q_{\text{a}} + Q_{\text{c}} \tag{32}$$

The cooling Coefficient of Performance (COP) in summer is:

$$\text{COP}_{\text{C}} = \frac{Q_{\text{e}}}{Q_{\text{g}} + W_{\text{sp}}} \tag{33}$$

$$W_{\text{sp}} = m_{20}(\text{h}_{21} - \text{h}_{20}) \tag{34}$$

The heating COP of the AHP in winter can be defined as:

$$\mathrm{COP_H} = \frac{Q_c + Q_a}{Q_g + W_{sp}} \tag{35}$$

**2.3 Organic Rankine cycle system**

The ORC is adopted to improve the electricity production. The ORC model incorporates two fundamental thermodynamic assumptions: the condenser outlet is saturated liquid, and the evaporator outlet is superheated steam. The organic working fluid is R245fa, which is an environmentally friendly fluid with a relatively low global warming potential and excellent thermodynamic performance [42]. The superheat temperature of the ORC system is 5 K, while the evaporating temperature is determined by the PEMFC temperature. The pump and turbine efficiencies are 0.85 and 0.85, respectively. The condensation temperatures are 298.15 K in summer and 283.15 K in winter. In the evaporator (Process 15→12 in Figure 1), the organic working fluid undergoes constant-pressure heat absorption, while in the condenser (Process 13→14 in Figure 1), the R245fa undergoes isobaric condensation. Therefore, the pressure relationship can be expressed as:

$$P_{15} = P_{12} \tag{36}$$

$$P_{14} = P_{13} \tag{37}$$

The isentropic efficiency of the turbine and the pump, $\eta_{\mathrm{tur}}$ and $\eta_{\mathrm{pump}}$, can be defined as follows:

$$\eta_{\mathrm{tur}} = \frac{h_{12} - h_{13}}{h_{12} - h_{13\mathrm{s}}} \tag{38}$$

$$\eta_{\mathrm{p}} = \frac{h_{15\mathrm{s}} - h_{14}}{h_{15} - h_{14}} \tag{39}$$

After thermal energy extraction in the evaporator, the R245fa undergoes adiabatic expansion in the turbine, resulting in power generation that can be defined as:

$$W_{\mathrm{ORC}} = m_{\mathrm{ORC}}(h_{12} - h_{13}) \tag{40}$$

The pump power consumption is:

$$W_{\mathrm{ORC,p}} = m_{\mathrm{ORC}}(h_{15} - h_{14}) \tag{41}$$

The net power and the efficiency of the ORC system can be defined as:

$$W_{\mathrm{ORC,net}} = W_{\mathrm{ORC}} - W_{\mathrm{ORC,p}} \tag{42}$$

$$\varepsilon_{\mathrm{ORC}} = \frac{W_{\mathrm{ORC,net}}}{Q_{\mathrm{st}}\zeta} \tag{43}$$

where $\zeta$ is the splitting ratio of waste heat entering the ORC system among all waste heat recovered by the PEMFC system. $\zeta$ = 0 represents that all the waste heat is delivered to the AHP, while $\zeta$ = 1 represents that all the waste heat is delivered to the ORC system.

Exergy analysis provides a quantitative benchmark for assessing the energy quality of a system [43]. In this study, the standard environmental pressure and temperature are 101 kPa and 298.15 K, respectively.

The physical exergy $ex_{ph}$ quantifies the maximum theoretical work obtainable when a system reaches thermodynamic equilibrium with its environment through temperature and pressure differentials, while the chemical exergy $ex_{ch}$ reflects the energy quality contained in the fuel due to the difference in chemical composition. These can be calculated as:

$$EX_{fuel} = m\left(ex_{ph} + ex_{ch}\right) \tag{44}$$

$$ex_{ph} = \left(h - h_0\right) - T_0\left(s - s_0\right) \tag{45}$$

$$ex_{ch} = \sum \chi_n ex_{ch}^{n} + RT_0 \sum \chi_n \ln x_n \tag{46}$$

The heat exergy $EX_H$ [44] and the cooling exergy $EX_C$ [38] of the system can be calculated as：

$$EX_H = \left(1 - \frac{T_0}{T_H}\right)Q \tag{47}$$

$$EX_C = \left(\frac{T_0}{T_C} - 1\right)Q_e \tag{48}$$

where $T_H$ and $T_C$ are the heating temperature and the cooling temperature, respectively. The exergy efficiency and the total net electrical power of the system can be defined as Eqs. (49) and (50).

$$\eta_{ex} = \frac{EX_{elec} + EX_H + EX_C}{EX_{fuel}} \tag{49}$$

$$W = W_{ORC,net} + W_{st,net} \tag{50}$$

The total energy of the inlet hydrogen $\hbar_{H_2}$ and the electrical efficiency $\eta$ are defined as Eqs. (51) and (52), respectively.

$$\hbar_{H_2} = \frac{n_{H_2}}{\beta_{H_2}} M_{H_2} HHV \tag{51}$$

$$\eta = \frac{W}{\hbar_{H_2}} \tag{52}$$

The thermal-to-electricity ratio of the system $\theta_{H,elec}$ is the ratio of the heating capacity to the electricity generated by the system, while the cooling-to-electricity ratio $\theta_{C,elec}$ is the ratio of the cooling capacity to the electricity generated by the system. They can be defined as:

$$\theta_{H,elec} = \frac{Q_H}{W} \tag{53}$$

$$\theta_{C,elec} = \frac{Q_C}{W} \tag{54}$$

## 2.4 System analysis

The economic viability and environmental friendliness are key factors in determining its feasibility and potential for widespread adoption. A comprehensive economic analysis considering environmental friendliness is conducted to analyze the competitiveness of the system[45]. Assuming that the system operates 8,000 hours a year, half of the time in summer mode, and half of the time in winter mode. The system is expected to have a lifespan of 20 years. The levelized cost of exergy (LCOE), annual total cost (ATC), net present value (NPV), capital recovery factor (CRF), dynamic payback period (DPP), and net cash flow (NCF) are important parameters for economic analysis, and they are analyzed in this study [46].

The ATC is determined by Eq. (55). The maintenance coefficient is denoted by $\omega_{\mathrm{m}}$, while $\tau$ represents annual operational hours and $Z_{\mathrm{k}}$ is the cost of all the subcomponents in the PEMFC-CCHP system ($). The financial analysis incorporates an interest rate $i$ over the $\vartheta$-year service life.

$$ATC = 3600 C_{\mathrm{H_2}} m_{\mathrm{H_2}} \tau + \sum Z_{\mathrm{k}} (CRF + \omega_{\mathrm{m}}) \tag{55}$$

$$CRF = \frac{i(i+1)^{\vartheta}}{(i+1)^{\vartheta} - 1} \tag{56}$$

The total cost of the initial equipment investment can be calculated as:

$$Z = Z_{\mathrm{PEMFC}} + Z_{\mathrm{ORC}} + Z_{\mathrm{AHP}} \tag{57}$$

where $Z_{\mathrm{PEMFC}}$, $Z_{\mathrm{ORC}}$, and $Z_{\mathrm{AHP}}$ represent the initial investment of the PEMFC, ORC, and AHP systems, respectively. The net cash flow can be obtained by subtracting annual costs from annual revenues. In comparison to a traditional system, the proposed system generates zero $CO_2$ emissions during operation, making it essential to incorporate carbon mitigation benefits in the economic evaluation. The $NCF$ can be calculated as

$$NCF = C_{\mathrm{H}} \Gamma_{\mathrm{H}} \tau_{\mathrm{H}} + C_{\mathrm{C}} \Gamma_{\mathrm{C}} \tau_{\mathrm{C}} + C_{\mathrm{elec}} \Gamma_{\mathrm{elec}} \tau + C_{\mathrm{ct}} - Z(CRF + \gamma_{\mathrm{m}}) - 3600 C_{\mathrm{H_2}} m_{\mathrm{H_2}} \tau \tag{58}$$

$$C_{\mathrm{ct}} = (\Upsilon_{\mathrm{H}} \tau_{\mathrm{H}} + \Upsilon_{\mathrm{C}} \tau_{\mathrm{C}} + \Upsilon_{\mathrm{elec}} \tau_{\mathrm{elec}}) \mathbb{N} \tag{59}$$

where $C_{\mathrm{H}}$, $C_{\mathrm{C}}$, and $C_{\mathrm{elec}}$ represent the heating cost, cooling cost, and electricity cost ($ kWh$^{-1}$), respectively; $\Gamma_{\mathrm{H}}$, $\Gamma_{\mathrm{C}}$, $\Gamma_{\mathrm{elec}}$ represents the thermal energy output, cooling energy output, and electrical energy output by the system (kWh); $\tau_{\mathrm{H}}$ and $\tau_{\mathrm{C}}$ represents the operation time in winter mode and summer modes (hours); $C_{\mathrm{H_2}}$ is the hydrogen price ($ kg$^{-1}$) and $m_{\mathrm{H_2}}$ is the mass flow rate of the hydrogen (kg s$^{-1}$). $\mathbb{N}$ is the carbon tax, and its value is set to 0.012 $ kg$^{-1}$ according to Ref. [47]. $\Upsilon_{\mathrm{H}}$, $\Upsilon_{\mathrm{C}}$, and $\Upsilon_{\mathrm{elec}}$ are the $CO_2$ emissions of the traditional heating, cooling, and electricity supply, and the values are 0.3978 kg $CO_2$ kWh$^{-1}$ [48], 0.5789 kg $CO_2$ kWh$^{-1}$ [49] and 0.8177 kg $CO_2$ kWh$^{-1}$ [48], respectively.

The levelized cost of exergy ($ kWh$^{-1}$) is defined as

$$LCOE = \frac{Z_{\mathrm{k}} (CRF + \omega_{\mathrm{m}}) + 3600 C_{\mathrm{H_2}} m_{\mathrm{H_2}} \tau}{W_{\mathrm{elec}} \tau + EX_{\mathrm{H}} \tau_{\mathrm{H}} + EX_{\mathrm{C}} \tau_{\mathrm{C}}} \tag{60}$$

and the net present value is

$$NPV = -\sum Z_{\mathrm{k}} + \sum_{\mathbb{Z}=1}^{\vartheta} \frac{NCF}{(1+i)^{\mathbb{Z}}} \tag{61}$$

The dynamic payback period represents the payback time in present value terms, which assesses the time required for a project to recover its initial investment. It is defined as

$$DPP = X + \frac{Y}{Z} \tag{62}$$

where $X$ represents the time when cash flow is negative, $Y$ is the absolute discounted cumulative cash flow value of the year when the cumulative present value becomes positive for the first time, and $Z$ is the first positive discounted cash flow.

The global sensitivity index is employed to quantify the influence of individual parameter variations on the economic performance of the system, which can be defined as,

$$S_{\mathrm{span}} = \frac{(\Theta_{\mathrm{max}} - \Theta_{\mathrm{min}}) / \Theta_0}{(\Lambda_{\mathrm{max}} - \Lambda_{\mathrm{min}}) / \Lambda_0} \tag{63}$$

where $\Theta$ is the dependent variable, and $\Lambda$ is the independent variable.

### 2.5 Model validation

The model of the PEMFC is verified with the data from the experiment to guarantee computational accuracy. The membrane was from Gore with a thickness of 12 μm, and the GDL was from TORAY with a thickness of 55 μm. The Pt loading was 0.1 mg $cm^{-2}$ for the anode and 0.4 mg $cm^{-2}$ for the cathode. The purity of hydrogen was 99.999%. The operating temperature of the PEMFC was 333.15 K, the pressure was 1 bar, and the relative humidity was 80%. A nominal activation area of 200 $cm^2$ was implemented for all unit cells in the assembly. The validation results are shown in Figure 2a, where the simulation results demonstrate excellent agreement with experimental measurements, exhibiting a maximum relative error of merely 2.6%, indicating that the model is accurate and can be used for the subsequent analysis.

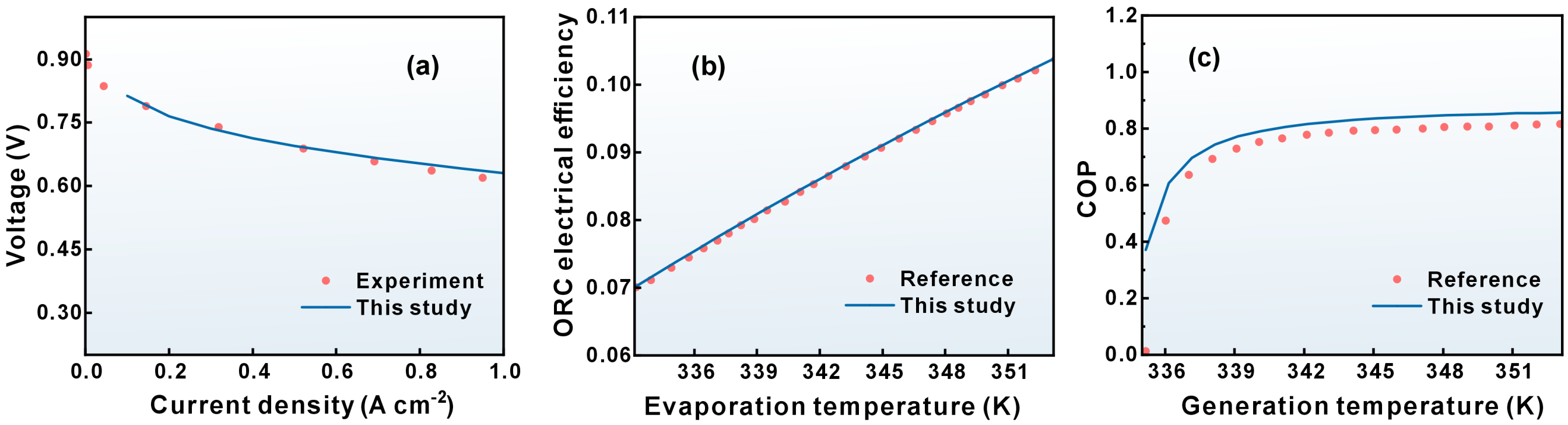

Figure 2: (a) Validation of PEMFC polarization curves; (b) Validation of ORC electrical efficiency based on Ref. [32]; (c) Comparison of the COP of the AHP between Ref. [50] and the simulation results.

The ORC system model is validated against Ref. [32]. The working fluid of the ORC system is R245fa, and the condensing temperature is 298.15 K with a superheat degree of 5 K (Figure 2b). The evaporation temperature of the system varies from 333.15 to 353.15 K, and the isentropic efficiency of the ORC fluid pump and the turbine is 0.85 and 0.85, respectively. The ORC electrical efficiency under different evaporation temperatures is shown in Figure 2b. The maximum difference between the simulation result and the reference result is 0.69 %. Hence, the accuracy of the ORC model is reliable.

A comparison between the simulation result of the AHP and Ref. [50] is performed to confirm the accuracy of the model. In the AHP system, lithium bromide serves as the absorbent, and water acts as the refrigerant; the effectiveness of the heat exchanger is 0.707. The condenser temperature is 303.15 K, the evaporator temperature is 278.15 K, and the absorber temperature is 303.15 K. The generation temperature of the system varies from 335.15 K to 353.15 K, and the COP under different generation temperatures is shown in Figure 2c. The results demonstrate that the maximum deviation is 4% between the simulation result and the reference result. Hence, the computational model can achieve acceptable accuracy.

## 3 Results and discussion

### 3.1 Energy flow in CCHP system

To quantitatively evaluate the energy conversion effectiveness of the proposed system and the synergistic effect between PEMFC, ORC, and AHP, Figure 3 quantifies the energy distribution characteristics of the system, revealing the energy flow of key components under seasonal operating conditions. In summer, the AHP provides cooling capacity, and the PEMFC and ORC systems provide electricity for the residential demands. The evaporation temperature of the ORC is decided by the PEMFC stack operating temperature. Affected by the ambient temperature, the ORC condensation temperature is set to 298.15 K. The generation temperature of the AHP decided by the stack temperature, the condensation temperature is 303.15 K, the absorption temperature is 298.15 K, and the evaporation temperature is 276.15 K. In winter, the AHP provides heating capacity, and the PEMFC and ORC systems provide electricity. The evaporation temperature of the ORC remains unchanged, and the condensation temperature is 283.15 K. The condensation temperature of the AHP is 325.15 K, the absorption temperature is 298.15 K, and the evaporation temperature is 283.15 K. The power consumption of the solution pump is not shown in the figure due to its negligible magnitude (within 0.01% of the total power). Nevertheless, this power consumption is fully accounted for in all quantitative analyses and results presented hereafter.

Energy flow diagrams provide a systematic visualization of the energy conversion processes within the PEMFC-CCHP system. Figure 3a shows the energy flow in summer mode, taking the energy of the input hydrogen as the reference value (100%), which corresponds specifically to 3544 W. Hydrogen enters the PEMFC stack for electrochemical reaction, and approximately 50.17% of the energy is converted to electricity, of which 3% enters the air compressor and 1671W for electrical power output. The waste heat from the stack and air compressors is recovered by cooling water and supplied to AHP and ORC, respectively. The splitting ratio $\zeta$ is 0.5 (i.e., 50% of the heat enters the ORC, and 50% enters the AHP).

Moreover, the ORC system supplies an additional 78 W of electricity, and the AHP produces 647 W of cooling capacity. Similarly, in the winter mode (Figure 3b), the PEMFC system maintains an equivalent electrical power output to that of the summer mode, while the ORC subsystem contributes an additional 98 W of electricity generation. Concurrently, the system delivers 1330 W of heating capacity to meet the residential heating demand. In summer mode, the system converts 49.34% of the energy input into net electrical power and 18.26% into cooling capacity. In winter mode, 49.93% is converted into net electrical power and 37.53% into heating capacity. The ORC can improve electrical efficiency by 2.19 percentage points in summer and 2.78 percentage points in winter. Moreover, the flexible energy supply is achieved by adjusting the splitting ratio of the waste heat, and the decoupling of electrical power output from heating/cooling can be realized.

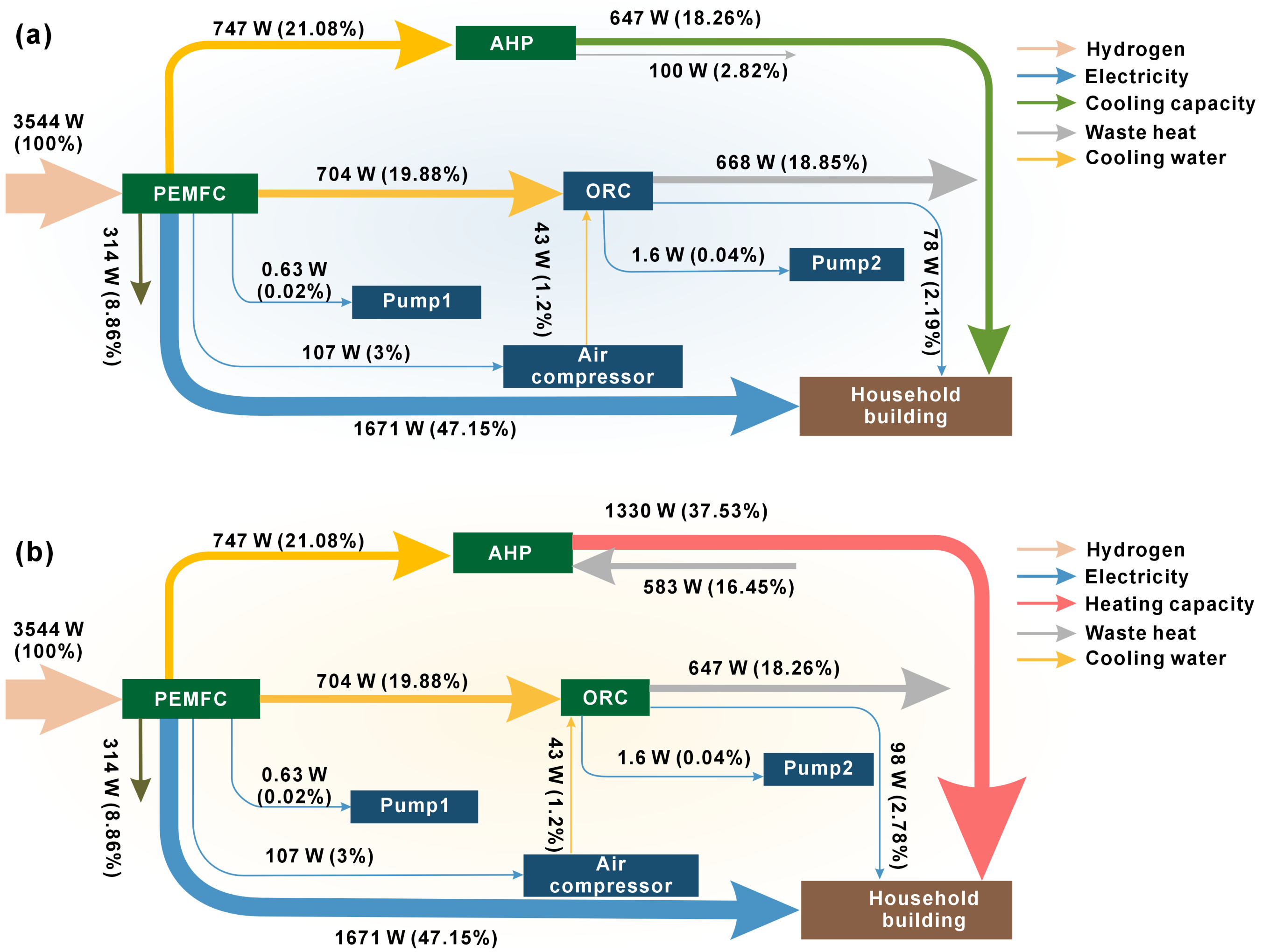

Figure 3: Energy flow diagrams of the PEMFC-CCHP system: (a) in summer mode, (b) in winter mode. Here, the PEMFC temperature is 353.15 K, the current density is 0.4 A $cm^{-2}$, and the splitting ratio is 0.5.

The PEMFC-CCHP system can decouple the electrical power supply and cooling/heating capacity. While meeting the basic cooling/heating demands, the remaining waste heat is assigned to the ORC system for electricity production. Additionally, key operational parameters of the stack, including current density and temperature, significantly impact electrical power output. As illustrated in Figure 4a, elevated current density leads to improved power generation capacity in the system. A higher splitting ratio indicates that more waste heat is utilized in the ORC for electricity production. Therefore, at a given current density, an increase in the splitting ratio will increase electrical power output. Higher temperatures reduce the activation losses in the PEMFC stack, thereby boosting electrical efficiency and consequently increasing electrical power output (see Figure 4b). The proportion of waste heat entering the ORC system and the PEMFC temperature can significantly affect the electrical power output. When the splitting ratio changes from 0 to 1, the electrical output power changes from 1661 to 1849 W (see Figure 4b).

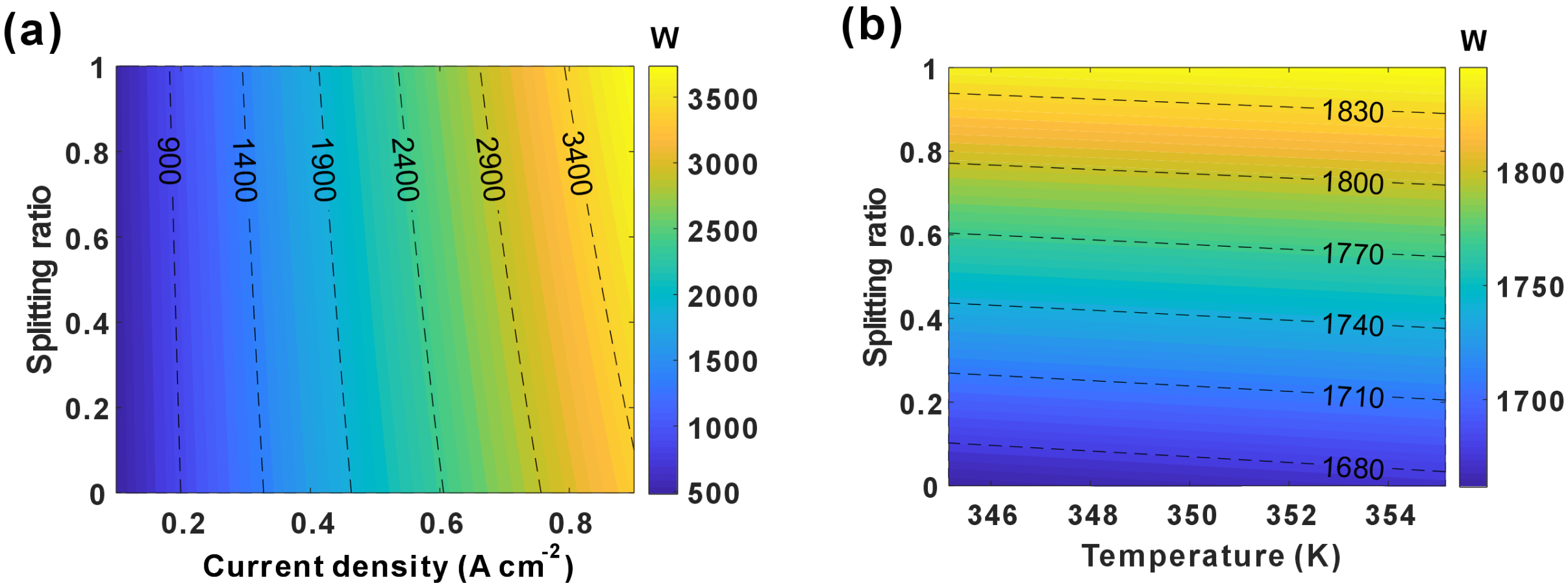

Figure 4: Effect of the splitting ratio $\zeta$ on the electrical power output: (a) electrical power output under different current densities (the PEMFC temperature is 353.15 K), (b) electrical power output under different PEMFC temperatures (the current density is 0.4 A cm$^{-2}$).

### 3.2 Summer mode

A systematic analysis is conducted to examine how key variables impact the operational states of the system. Figures 5 and 6 examine the influence of key operational parameters on the thermodynamic characteristics of the system in summer mode. As PEMFC temperature changes from 345.15 to 355.15 K, the cooling COP increases from 0.8604 to 0.8667, but the cooling-to-electricity ratio decreases from 0.379 to 0.368 (see Figure 5a). It has been demonstrated that elevated generation temperature enhances the concentration difference between the lithium bromide solution at the generator and the absorber, which improves the absorption capacity and also the heat utilization efficiency of the generator, thus increasing the cooling COP. Elevating the operating temperature of the PEMFC enhances both its electrical conversion efficiency and the ORC efficiency accordingly. The exergy efficiency and electrical efficiency increase (Figure 5b), thereby boosting the electrical power output (Figure 5c). However, the elevated electrical efficiency reduces the available waste heat, leading to a decline in cooling capacity as PEMFC temperature increases (Figure 5c).

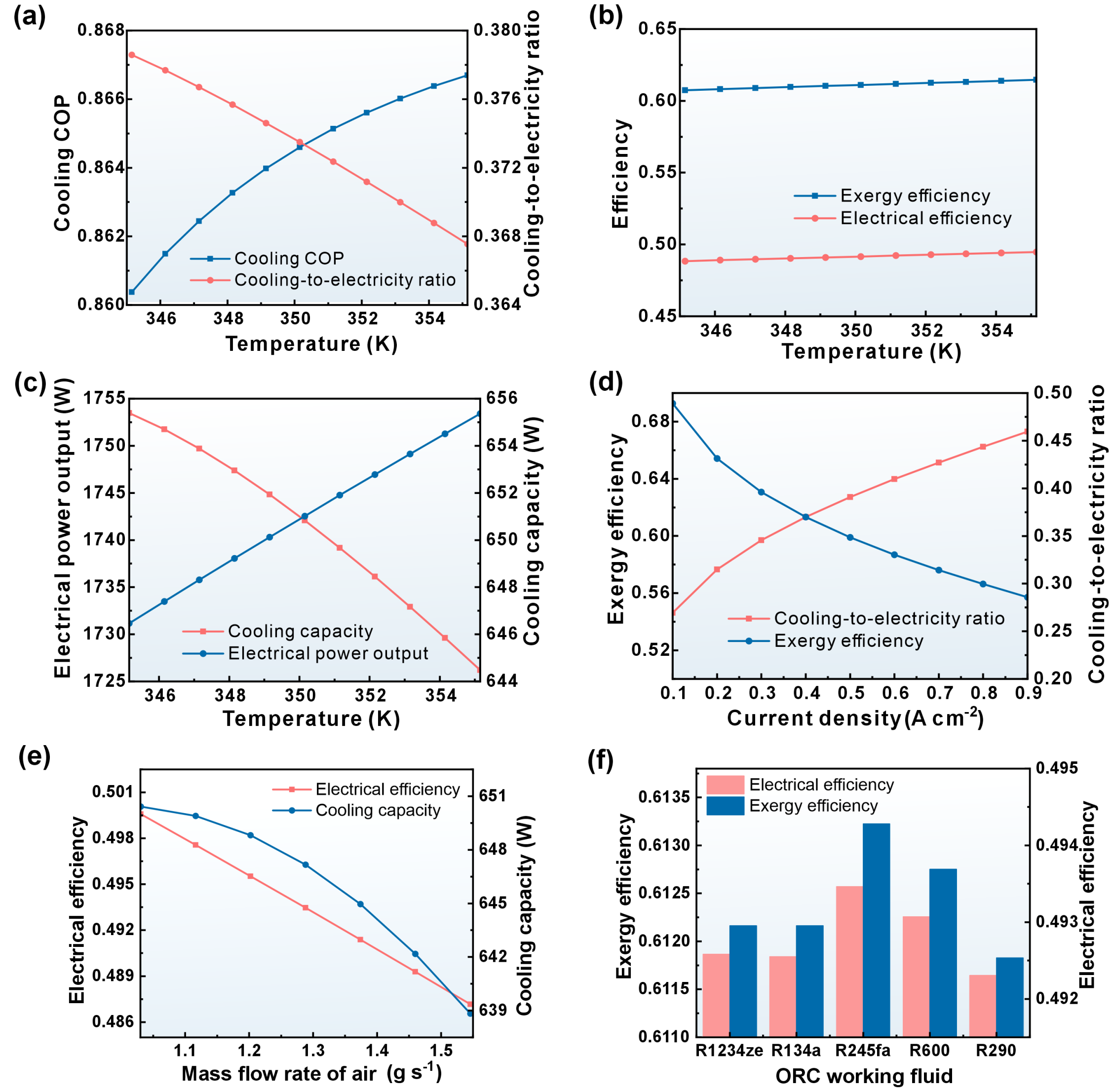

Figure 5: Influence of temperature and current density of PEMFC in summer mode: (a) cooling COP and cooling-to-electricity ratio versus temperature; (b) electrical efficiency and exergy efficiency versus temperature; (c) cooling capacity and electrical power output versus temperature; (d) cooling-to-electricity ratio and exergy efficiency versus current density; (e) electrical efficiency and cooling capacity versus mass flow rate of air; (f) electrical efficiency and exergy efficiency versus ORC working fluid.

Increasing the current density from 0.1 to 0.9 A $cm^{-2}$ leads to more irreversible losses within the PEMFCs, causing a corresponding reduction in system exergy efficiency from 0.69 to 0.55 (Figure 5d). Concurrently, the irreversible losses enter the AHP as waste heat, which enhances the cooling capacity and leads to an increase in the cooling-to-electricity ratio of the system (Figure 5d). When the mass flow rate of air varies from 1.03 to 1.55 g $s^{-1}$, the system cooling capacity decreases by 11.58 W (Figure 5e), and the electrical efficiency drops by 0.012. This is because the increased excess air at the cathode leads to more waste heat being carried away by the cathode exhaust, reducing the heat available for power generation and cooling capacity. The performance of the system employing five different organic working fluids is shown in Figure 5f, and the results indicate that R245fa achieves the highest exergy efficiency of 0.61, followed by R600; R134a and R1234ze exhibit similar thermodynamic performance. Furthermore, the variation trend of the electrical efficiency is found to be similar to that of the exergy efficiency.

The modulation of the splitting ratio and critical operational parameters enables precise regulation of the cooling capacity, thereby adapting to diverse demands. An increase in the current density results in enhanced irreversible losses, which consequently leads to an enhancement in the cooling capacity. Meanwhile, as the splitting ratio rises, the cooling water flow to the ORC branch increases, and the cooling capacity decreases. When the current density is fixed at 0.4 A cm$^{-2}$ and the splitting ratio increases from 0 to 0.5, the cooling capacity of the system varies from 1294 to 647 W (see Figure 6a). As current density increases from 0.1 to 0.9 A cm$^{-2}$ ($\zeta$ = 0), there is a significant rise in cooling capacity from 268 to 3263 W, as shown in Figure 6a. The current density affects the cooling capacity more significantly than the PEMFC temperature. With the increase in temperature from 345.15 to 355.15 K, the cooling capacity reduces from 1311 to 1289 W ($\zeta$ = 0), as shown in Figure 6b.

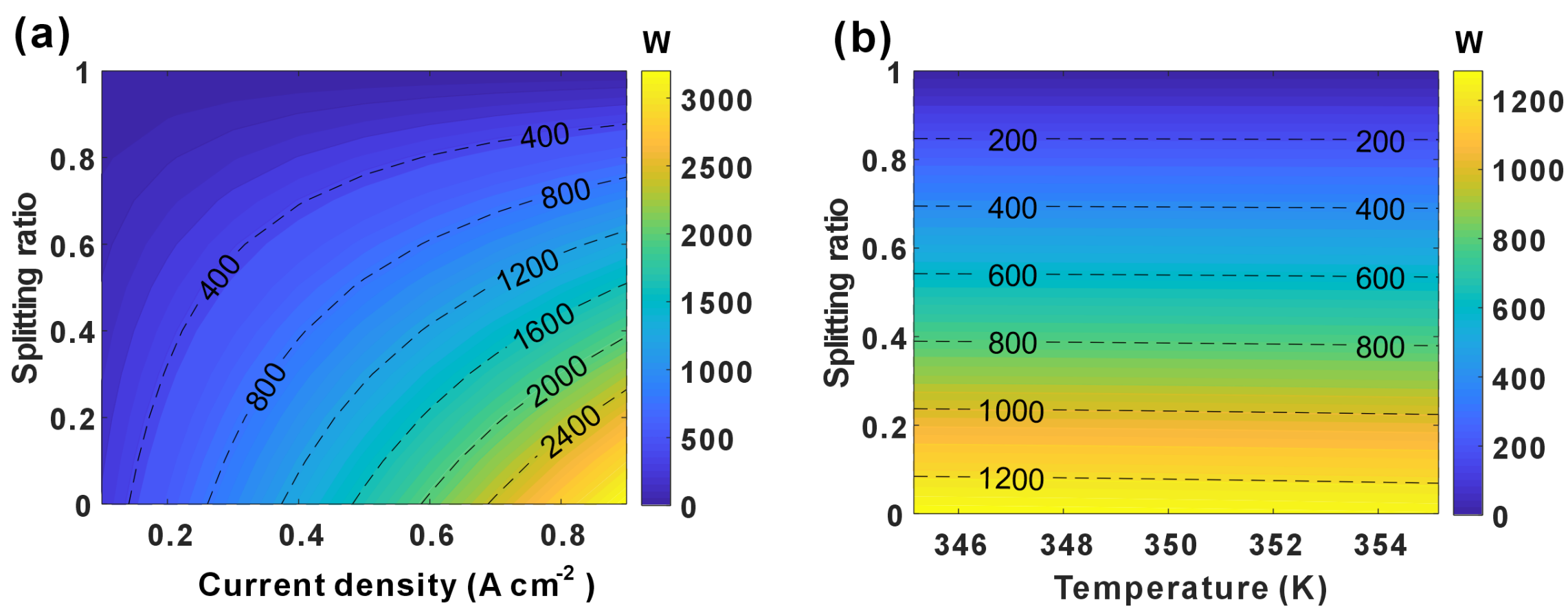

Figure 6: Effect of the splitting ratio on the cooling capacity: (a) cooling capacity under different current densities at the PEMFC temperature of 353.15 K; (b) cooling capacity under different PEMFC temperatures at the current density of 0.4 A cm$^{-2}$.

### 3.3 Winter mode

In winter, the influences of operational parameters on energy utilization are shown in Figures 7 and 8. Elevating the PEMFC temperature from 345.15 to 355.15 K results in a significant enhancement in heating COP from 1.05 to 1.79 (see Figure 7a), and concurrent improvements in both exergy efficiency and ORC efficiency (see Figure 7b). Meanwhile, the heating capacity increases from 802 to 1333 W, while the electrical power output rises slightly from 1753 to 1773 W (see Figure 7c). The more pronounced growth in heating capacity relative to electrical power output results in an ongoing rise in the thermal-to-electricity ratio from 0.46 to 0.75 with increasing PEMFC temperature (see Figure 7a). The increased generation temperature causes the system to produce a higher concentration of solution at the generator outlet, which intensifies the evaporation process of the refrigerant, and more refrigerant enters the condenser, with increased heat release in the condenser. In addition, the enhanced concentration gradient between the generator and the absorber intensifies the absorption process in the absorber, and the heat release in the absorber increases, consequently causing an increase in heating capacity. Additionally, higher current density contributes to greater irreversible losses, lowering the exergy efficiency from 0.72 to 0.59 as the current density rises from 0.1 to 0.9 A cm$^{-2}$. Additionally, it enhances heat generation within the system, leading to an increased thermal-to-electricity ratio (see Figure 7d). An increase in the mass flow rate of air from 1.03 to 1.55 g s$^{-1}$ reduced the heating capacity by 23.81 W and the electrical efficiency by 0.013. The highest electrical efficiency (0.104) and power output (77.55 W) were achieved using R245fa as the ORC working fluid.

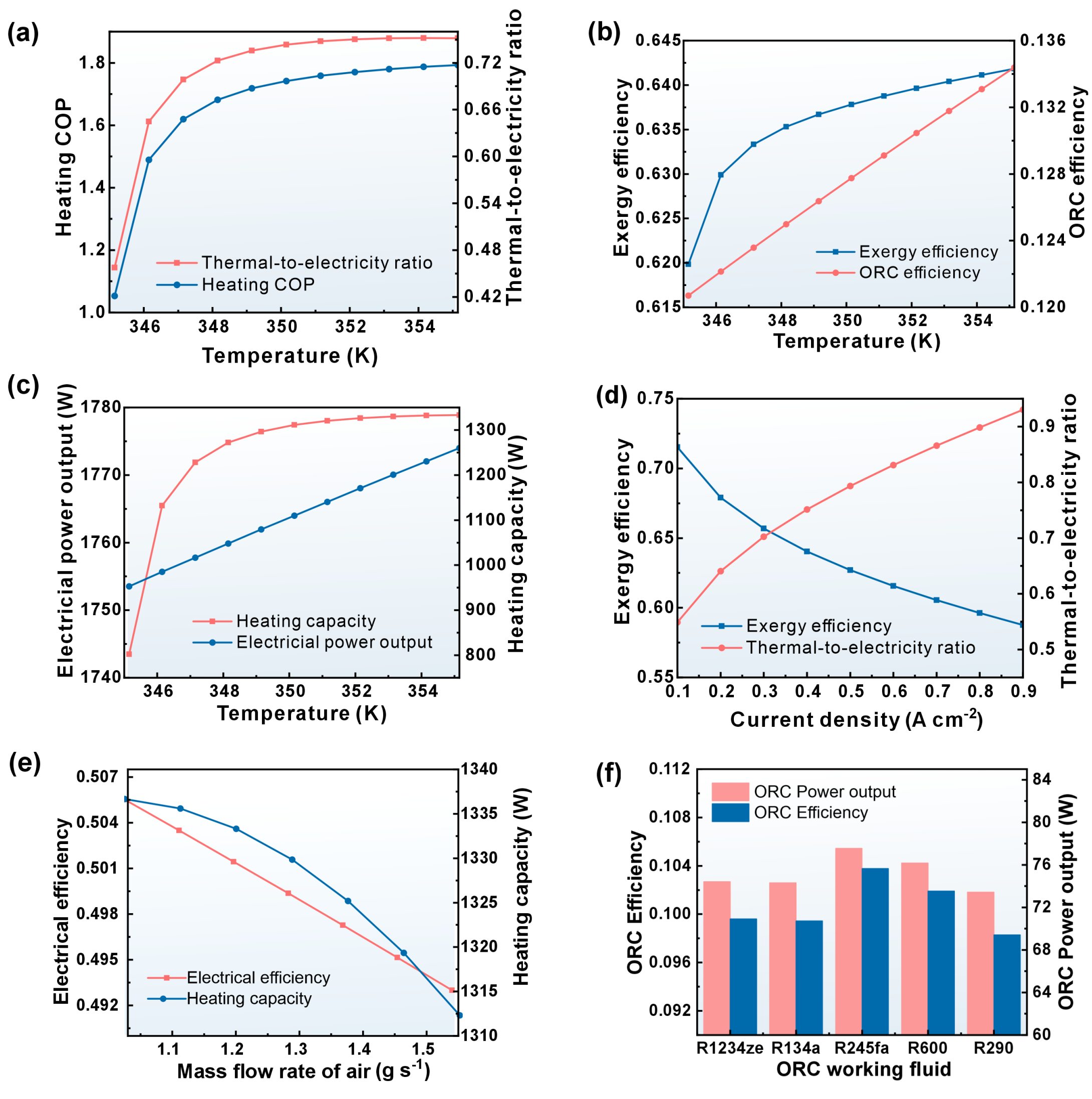

Figure 7: Influence of temperature and current density of PEMFC in winter mode: (a) heating COP and thermal-to-electricity ratio versus temperature; (b) exergy efficiency and ORC efficiency versus temperature; (c) heating capacity and electrical power output versus temperature; (d) thermal-to-electricity ratio and exergy efficiency versus current density; (e) electrical efficiency and heating capacity versus mass flow rate of air; (f) ORC Power output and ORC Efficiency versus ORC working fluid.

The heating capacity of the system in winter is directly affected by the splitting ratio and the operating parameters of the stack. By adjusting the splitting ratio, the flexible utilization of waste heat can be achieved, and the decoupling of electricity generation and heating generation is realized. Figure 8 illustrates the heating capacity under different splitting ratios. An elevated current density enhances the irreversible losses and leads to an increase in the heating capacity. With the increase in current density from 0.1 to 0.9 A cm$^{-2}$ ($\zeta$ = 0), there is a significant rise in heating capacity from 551 to 6707 W (see Figure 8a). In comparison, the effect of the PEMFC temperature is relatively moderate: raising the PEMFC temperature from 345.15 to 355.15 K results in a limited increase in heating capacity, from 1604 to 2667 W ($\zeta$ = 0), as shown in Figure 8b. Conversely, increasing the splitting ratio results in a greater diversion of cooling water to the ORC system, thereby reducing the available heating capacity. When the current density is fixed at 0.4 A cm$^{-2}$ and the splitting ratio increases from 0 to 0.5, the heating capacity of the system decreases from 2660 to 1330 W (Figure 8b).

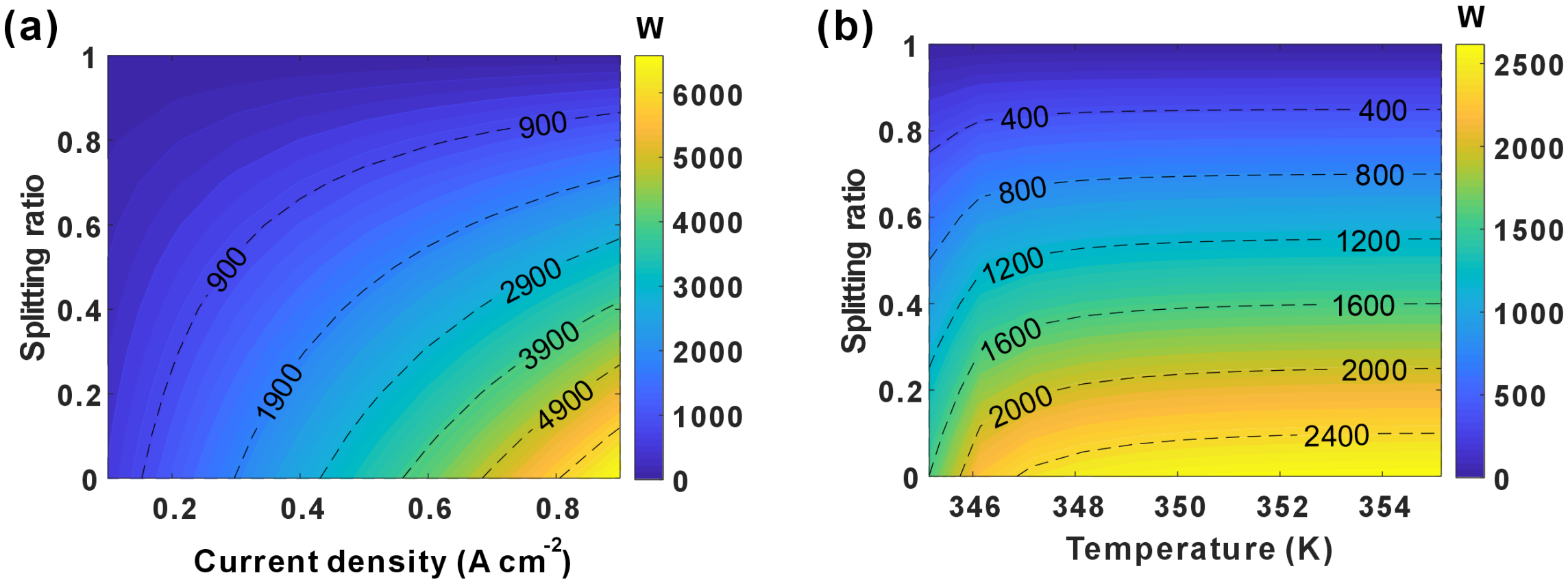

Figure 8: Effect of the splitting ratio on the heating capacity: (a) heating capacity under different current densities (PEMFC temperature is 353.15 K); (b) heating capacity under different PEMFC temperatures (current density is 0.4 A cm$^{-2}$).

### 3.4 Economic analysis

The price of hydrogen gas significantly impacts the economic performance of the system. The U.S. Department of Energy has set a target to lower the green hydrogen price to 2 \$ kg$^{-1}$ by 2026 and 1 \$ kg$^{-1}$ by 2031 [51]. In this study, the price range of hydrogen gas is set at 1–3 \$ kg$^{-1}$. The base electricity price, heating price, cooling price, and carbon tax are 0.136, 0.059, 0.1823, and 0.012 \$ kg$^{-1}$, respectively. As a key economic indicator, net present value (NPV) quantifies project viability by discounting all future cash flows to current value and subtracting the initial expenditure. The dynamic payback period (DPP) measures the time required for a project to recoup its initial investment. The levelized cost of exergy (LCOE) represents the average expenditure required to produce one unit of available energy, encompassing capital expenditures, operation and maintenance, and fuel costs. Net Cash Flow (NCF) refers to the difference between the actual cash inflows and the actual cash outflows during a specific period. Internal Rate of Return (IRR) is the discount rate that makes the NPV of an investment project equal to zero throughout its entire life cycle.

A combined analysis of NPV, DPP, LCOE, and IRR provides a comprehensive assessment of the economic viability and risk balance of the system. The tornado diagrams and global sensitivity coefficients are employed to systematically evaluate the impact of key economic parameters on system economic performance.

The LCOE and NCF exhibit complex dependencies on the current density and the splitting ratio, governed by the synergistic coupling of the electro-thermal-cooling multi-energy flows, as illustrated in Figure 9. The LCOE decreases monotonically with increasing current density, reducing by up to 48.36%–48.79% at lower current densities (0.1–0.5 A $cm^{-2}$) and narrowing to a reduction of 7.43%–7.55% at higher current densities (0.5–0.9 A $cm^{-2}$). Meanwhile, the NCF increases significantly with rising current density, with increases by 1314–2884 $ at lower current densities (0.1–0.5 A $cm^{-2}$) and 1080–2872 $ at higher current densities (0.5–0.9 A $cm^{-2}$), respectively.

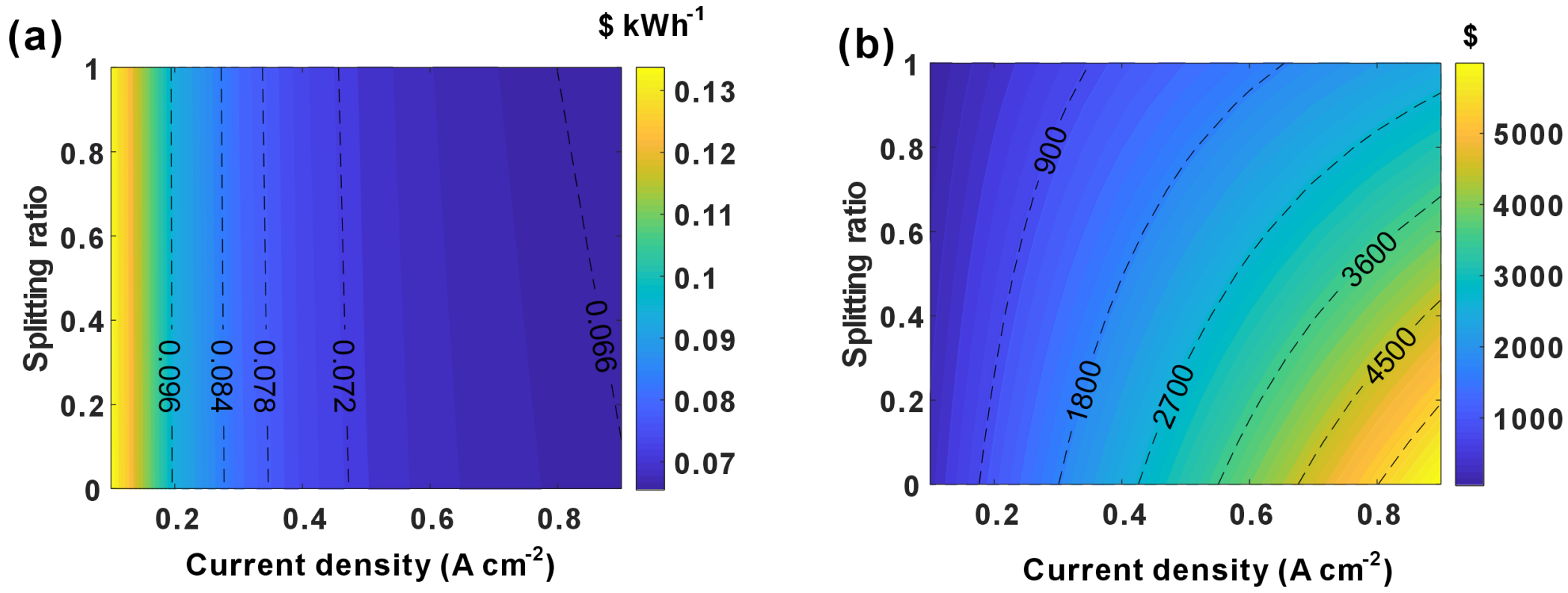

Figure 9: Impact of current density on system economic performance: (a) levelized cost of exergy (LCOE); (b) net cash flow (NCF).

Figure 10 illustrates how the PEMFC temperature and the splitting ratio affect the system economics. LCOE exhibits a pronounced decline with increasing the PEMFC temperature and the splitting ratio (see Figure 10a), reducing by up to 0.23%-2.6% at lower temperatures (345.15–350.15 K) and narrowing to a reduction of 0.21%–0.41% at higher temperatures (350.15–355.15 K). The results are primarily driven by two synergistic mechanisms: the activation polarization losses in the PEMFC stack decrease significantly with rising temperature; moreover, the thermoelectric conversion efficiency of the ORC system continues to improve with the increase of PEMFC temperature. The NCF shows a slight upward trend with increasing temperature, but decreases significantly with increasing splitting ratio (see Figure 11b). When the splitting ratio increases from 0 to 1, the NCF decreases sharply by 53.12%-57.73%, with higher temperatures mitigating the decline. However, temperature variations exhibit limited influence on this economic indicator, and the splitting ratio has a much stronger regulatory effect on the system economics than the PEMFC temperature.

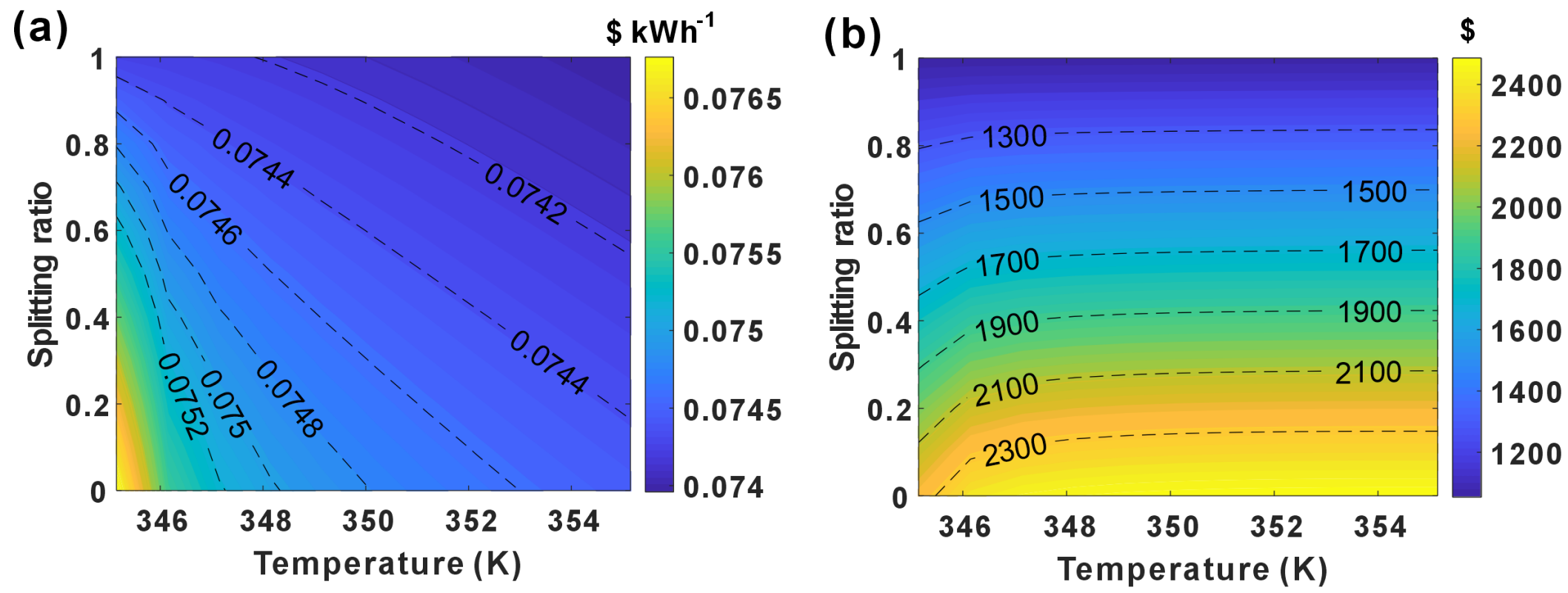

Figure 10: Impact of PEMFC temperature on system economic performance: (a) levelized cost of exergy (LCOE); (b) net cash flow (NCF).

The NPV under different hydrogen prices is shown in Figure 11. Within the system lifespan range, the NPV increases to a positive value, indicating acceptable economic performance. When the fuel price increases from 1 to 3 $ kg$^{-1}$, the dynamic payback period of the system changes from 2.44 to 16.94 years. The system electricity price has a considerably more pronounced impact on the dynamic payback period than either the cooling or heating price. A 20% increase in the electricity price reduces the dynamic payback period by 1.21 years, whereas a comparable 20% increase in the cooling and heating prices shortens the period by only 0.38 years and 0.26 years, respectively. In contrast, a 20% rise in the carbon tax extends the dynamic payback period by 0.16 years (Figure 11).

Due to the high volatility of hydrogen prices, as the hydrogen price rises from 1 to 3 $ kg$^{-1}$, the IRR decreases significantly from 43.87% to 5.31%. Similarly, when the electricity price, carbon tax, cooling price, and heating price vary by ±20% from their baseline values, the corresponding IRR ranges are 15.66%–35.49%, 24.87%–26.73%, 23.36%–28.22%, and 24.18%–27.41%, respectively (Table 1). The sensitivity analysis reveals that electricity price and hydrogen price are the most influential factors (Figure 12), with sensitivity coefficients relative to NPV of 2.54 and 1.91, and to IRR of 1.92 and 1.49, respectively, significantly higher than those of cooling price, heating price, and carbon tax. Thus, electricity and hydrogen prices are identified as the primary drivers of NPV and IRR (Table 2).

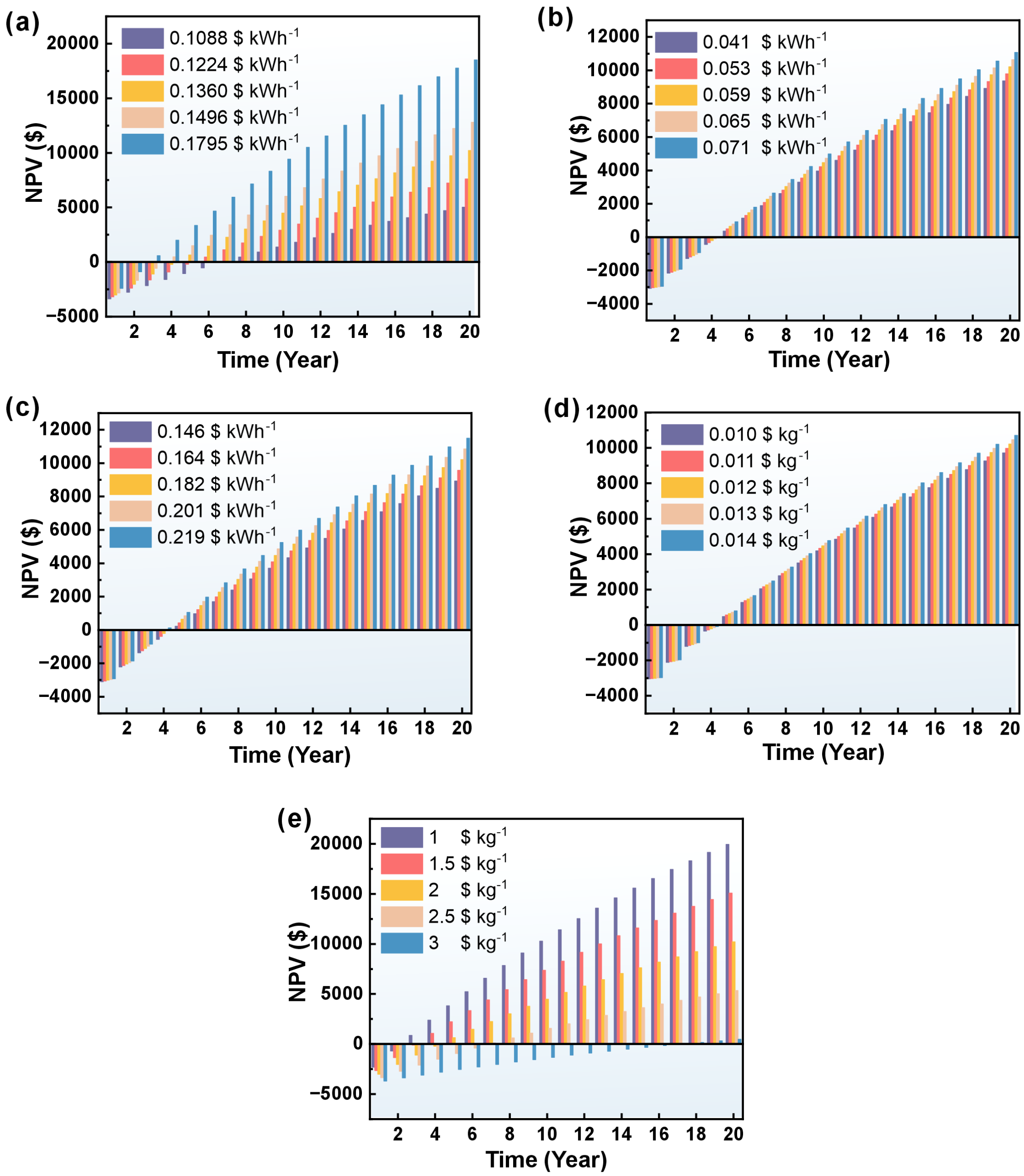

Figure 11: Impact of economic factors on NPV: (a) electricity prices, (b) heating price, (c) cooling price, (d) carbon tax, (e) hydrogen price.

Table 1: Impact of economic factors on IRR.

| Electricity price ($ kWh$^{-1}$) | IRR (%) | Heating price ($ kWh$^{-1}$) | IRR (%) | Cooling price ($ kWh$^{-1}$) | IRR (%) | Carbon tax ($ kg$^{-1}$) | IRR (%) | Hydrogen price ($ kg$^{-1}$) | IRR (%) |
|---|---|---|---|---|---|---|---|---|---|
| 0.109 | 15.66 | 0.047 | 24.18 | 0.15 | 23.36 | 0.001 | 24.87 | 1 | 43.87 |
| 0.122 | 20.83 | 0.053 | 25.00 | 0.16 | 24.59 | 0.011 | 25.34 | 1.5 | 34.90 |
| 0.136 | 25.80 | 0.059 | 25.80 | 0.18 | 25.80 | 0.012 | 25.80 | 2 | 25.80 |
| 0.150 | 30.67 | 0.065 | 26.61 | 0.20 | 27.01 | 0.013 | 26.27 | 2.5 | 16.32 |
| 0.180 | 35.49 | 0.071 | 27.41 | 0.22 | 28.22 | 0.014 | 26.73 | 3 | 5.31 |

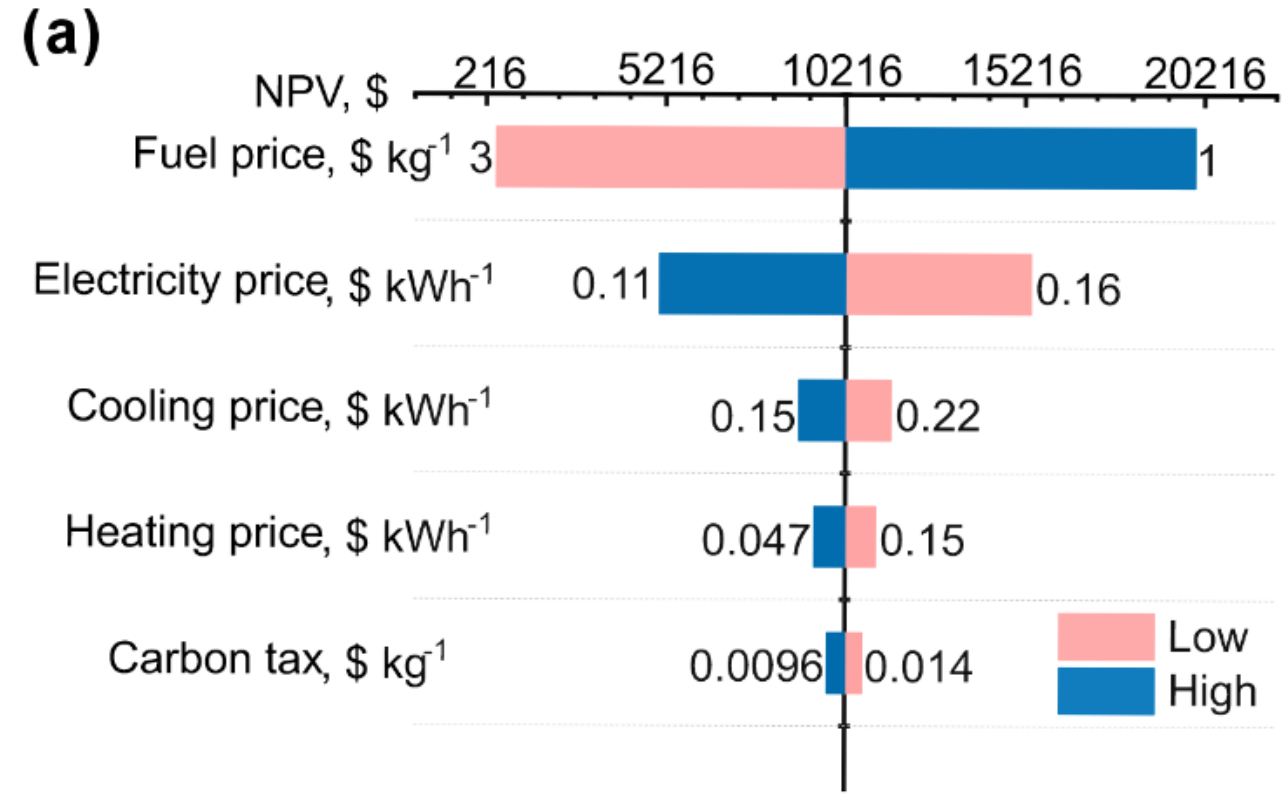

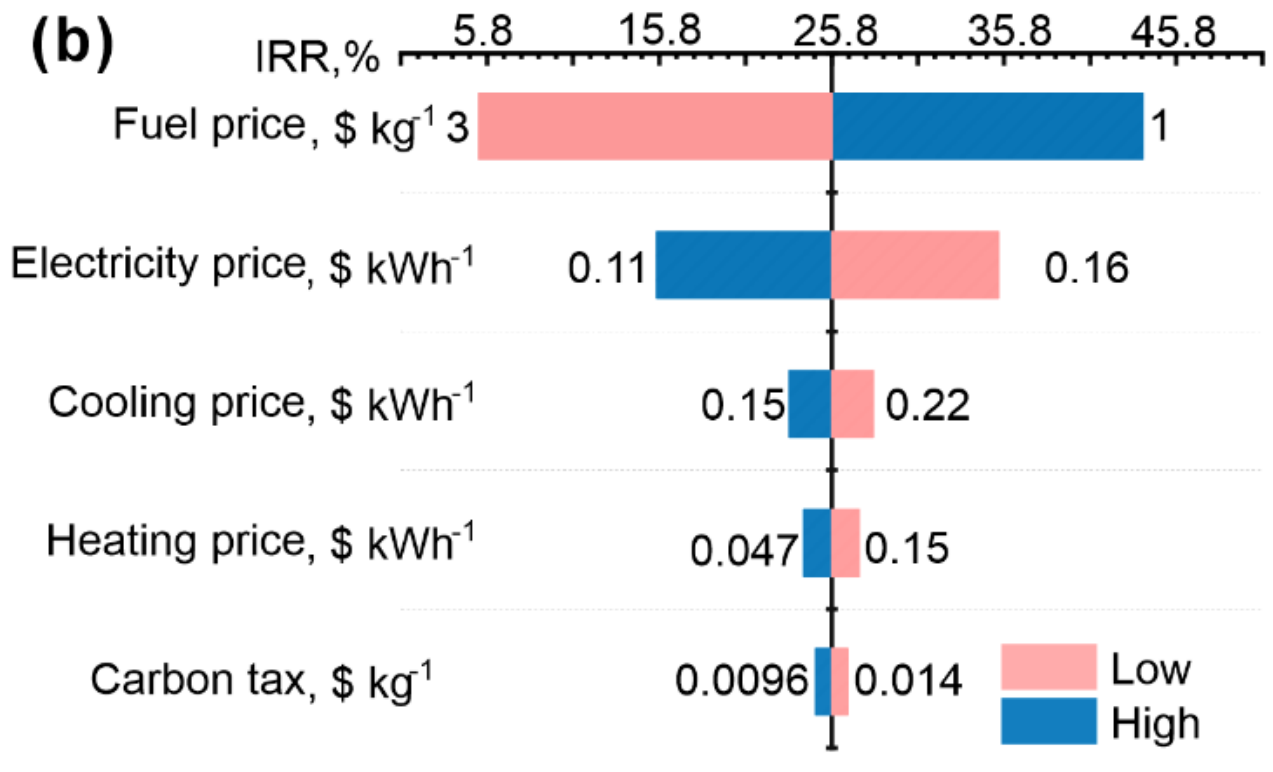

Figure 12: Tornado diagrams for (a) NPV sensitivity analysis, and (b) IRR sensitivity analysis.

Table 2: Global sensitivity coefficient.

| Global sensitivity coefficient | NPV | IRR |
| --- | --- | --- |
| Electricity price | 2.54 | 1.92 |
| Heating price | 0.42 | 0.31 |
| Cooling price | 0.63 | 0.47 |
| Hydrogen price | 1.91 | 1.49 |
| Carbon tax | 0.24 | 0.18 |

## 4. Conclusions

A PEMFC-CCHP system is proposed that recycles waste heat of PEMFC through a parallel arrangement of an ORC system and an AHP, which can decouple cooling, heating, and electrical power output to adapt to diverse residential energy demands. The energy flow is evaluated using energy flow diagrams both in summer and winter modes, and the effects of key parameters on thermodynamic and economic performance are discussed. The following conclusions can be drawn based on the analysis of the results:

(1) The organic Rankine cycle can improve electrical efficiency by 2.19 and 2.78 percentage points in summer and winter, respectively (at a splitting ratio of 0.5).

(2) The power output characteristics demonstrate strong dependence on both PEMFC operating parameters and splitting ratio. When the splitting ratio increases from 0 to 1, the electrical output power changes from 1661 to 1849 W. At a splitting ratio of 0, the cooling capacity increases markedly from 268 W to 3263 W, as the current density rises from 0.1 to 0.9 $A \cdot cm^{-2}$ at 353.15 K. In contrast, elevating the PEMFC temperature from 345.15 to 355.15 K, under a constant current density of 0.4 $A \cdot cm^{-2}$, leads to an increase in heating capacity from 1604 to 2667 W.

(3) The sensitivity analysis reveals that electricity and hydrogen prices are the primary drivers of NPV and IRR, with sensitivity coefficients relative to NPV of 2.54 and 1.91, and to IRR of 1.92 and 1.49, respectively—significantly higher than those of cooling price, heating price, and carbon tax.

(4) Reducing the hydrogen price is a vital route to enhancing the economy of the system. The dynamic payback period reduces from 16.94 to 2.44 years as the hydrogen price reduces from 3 to 1 $ $kg^{-1}$.

The PEMFC-CCHP system presented in this study is significant not only for both the engineering design and practical implementation of fuel cell-based trigeneration systems, but also helpful for the optimization of other application scenarios of hydrogen energy. Future work could focus on the dynamic characteristics and system optimization of the PEMFC-CCHP system, which in this study is examined under steady-state conditions. This could involve developing predictive models for cooling, heating, and electricity demand, and establishing a data-driven energy management strategy to optimize stack operating parameters and waste heat allocation. Ultimately, the aim would be to achieve coordinated optimization that balances dynamic load response, energy utilization efficiency, and occupant comfort holistically.

## Acknowledgements

This work was financially supported by the Department of Science and Technology of Inner Mongolia Autonomous Region (Grant No. 2022JBGS0027).